\documentclass[preprint,pre]{revtex4}
\usepackage[T1]{fontenc}
\usepackage[latin1]{inputenc}
\usepackage[dvips]{graphicx}
\usepackage{float}
\usepackage{amssymb}
\usepackage{amsmath}

\newcommand{\lav}{\left\langle}
\newcommand{\rav}{\right\rangle}

\restylefloat{figure}

\begin{document}
\title{Molecular Traffic Control in Porous Nanoparticles}
\author{Andreas Brzank\textsuperscript{1,2}}
\email{a.brzank@fz-juelich.de}
\author{Gunter Schütz\textsuperscript{1}}
\affiliation{\textsuperscript{1}Institut f\"ur Festk\"orperforschung, Forschungszentrum J\"ulich,
52425 J\"ulich, Germany\\
\textsuperscript{2}Fakultät für Physik und Geowissenschaften, Universität Leipzig, Abteilung Grenzflächenphysik,
Linnestrasse 5, D-04103 Leipzig, Germany }
\date{\today}

\begin{abstract}
We investigate the conditions for reactivity enhancement of catalytic
processes in porous solids by use of molecular traffic control (MTC) as a
function of reaction rate and grain size. With dynamic Monte-Carlo simulations
and continuous-time random walk theory applied to the low concentration regime
we obtain a quantitative description of the MTC effect for a network of intersecting
single-file channels in a wide range of grain parameters and for optimal external
operating conditions. The efficiency ratio (compared with a topologically and
structurally similar reference system without MTC) is inversely proportional
to the grain diameter. However, for small grains MTC leads to a reactivity
enhancement of up to approximately 30\% of the catalytic conversion $A\to B$
even for short intersecting channels. This suggests that MTC may significantly
enhance the efficiency of a catalytic process for small porous nanoparticles
with a suitably chosen binary channel topology.
\end{abstract}
\maketitle

\section{Introduction}

Zeolites are used for catalytic processes in a variety of applications, e.g.
cracking of large hydrocarbon molecules. In a number of zeolites diffusive
transport occurs along quasi-one-dimensional channels which do not allow guest
molecules to pass each other \cite{Baer01}. Due to mutual blockage of reactand
$A$ and product molecules $B$ under such {\it single-file conditions}
\cite{Karg92} the effective reactivity of a catalytic process $A\to B$ -- determined
by the residence time of molecules in the zeolite -- may be considerably reduced
as compared to the reactivity in the absence of single-file behaviour.
It has been suggested that the single-file effect may be circumvented by the so far
controversial concept of molecular traffic control (MTC) \cite{Dero80,Dero94}.
This notion rests on the assumption that reactands and product molecules resp.
may prefer spatially separated diffusion pathways and thus avoid mutual
suppression of self-diffusion inside the grain channels.

The necessary (but not sufficient) requirement for the MTC effect, a channel
selectivity of two different species of molecules, has been verified by
means of molecular dynamic (MD) simulations of two-component mixtures
in the zeolite ZSM-5 \cite{Snur97} and relaxation simulations of a
mixture of differently sized molecules (Xe and SF$_6$) in a bimodal
structure possessing dual-sized pores (Boggsite with 10-ring and 12-ring pores)
\cite{Clar00}. Also equilibrium Monte-Carlo simulations demonstrate that the
residence probability in different areas of the intracrystalline pore space
may be notably different for the two components of a binary mixture
\cite{Clar99} and thus provide further support for the notion
of channel selectivity in suitable bimodal channel topologies.

\begin{figure}
\centerline{
\includegraphics[width=6cm]{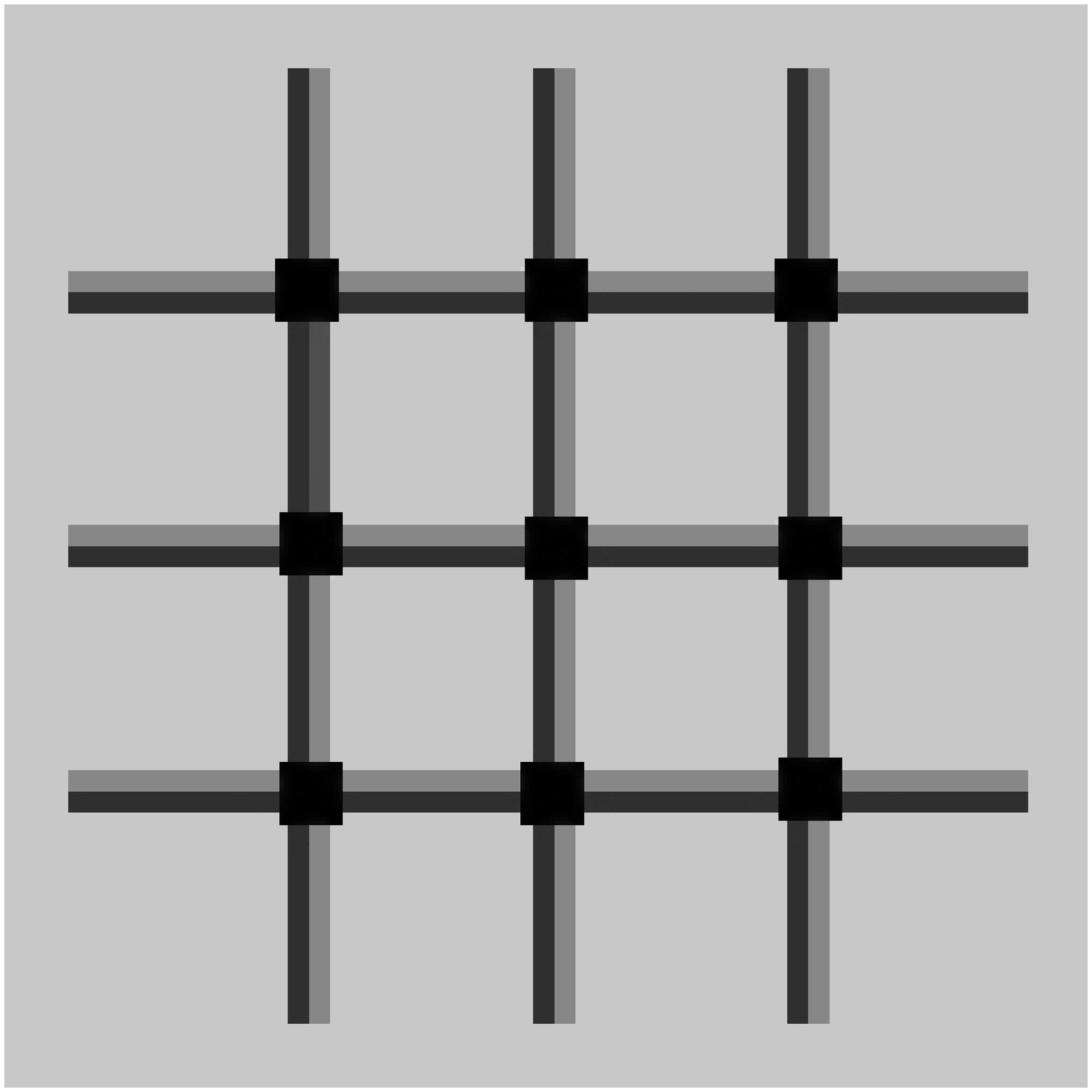}
\includegraphics[width=6cm]{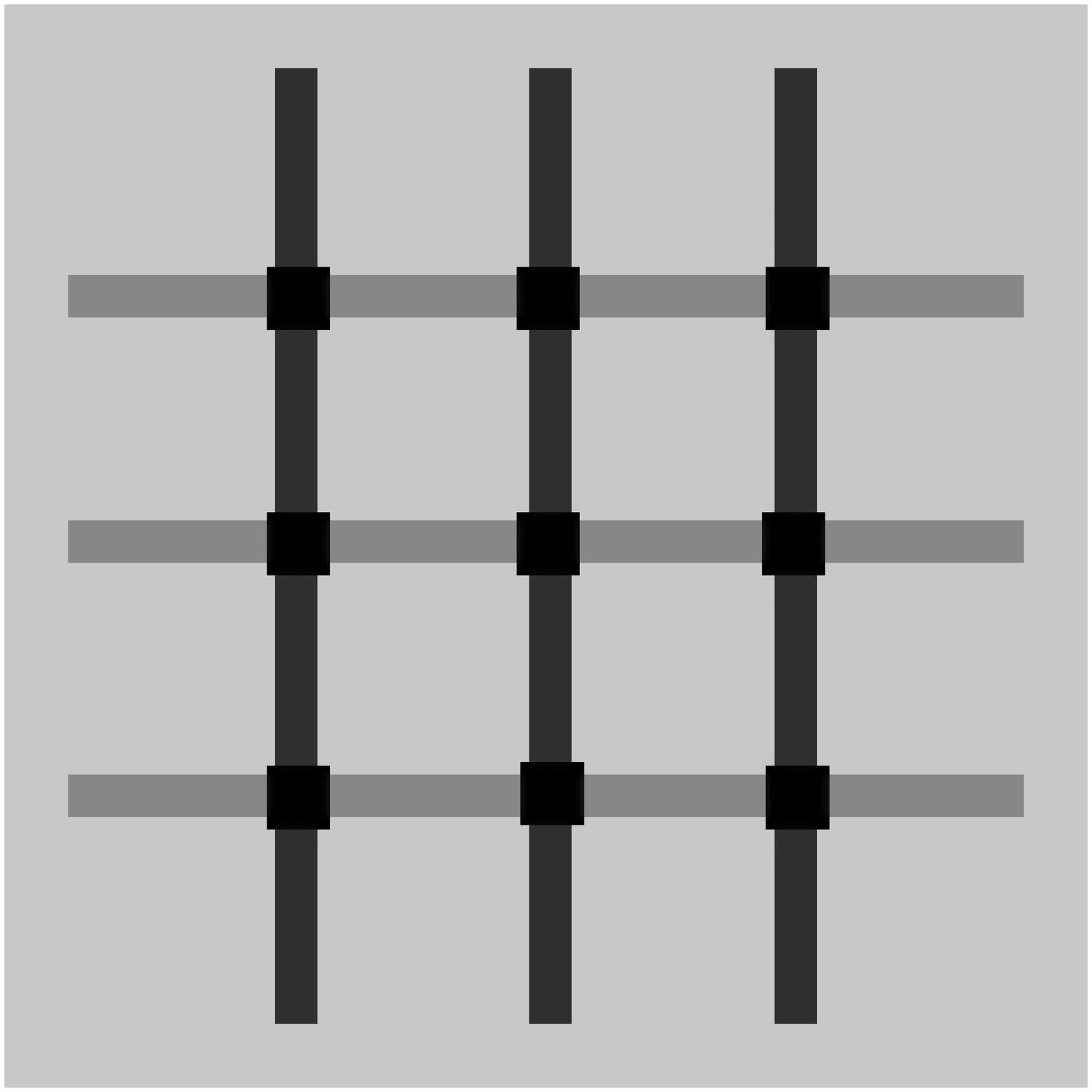}
}
\caption{REF system (left) with $N=3$ channels and MTC system
(right) of the the same size. In contrast to the REF case, where we allow both
types of particles ($A$ and $B$ particles) to enter any channel, in the MTC
system $A$ particles are carried through the vertical $\alpha$
channels whereas the $B$ particles diffuse along the horizontal $\beta$
channels. Black squares indicate catalytic sites where a catalytic
transformation $A\to B$ is allowed.}
\label{systemPics}
\end{figure}

Whether a MTC effect leading to reactivity enhancement actually takes place was
addressed by a series of dynamic Monte Carlo simulations (DMCS) of a stochastic
model system with a network of perpendicular sets of bimodal intersecting channels
and with catalytic sites located
at the intersecting pores (NBK model) \cite{Neug00,Karg00,Karg01}. The authors
of these studies found numerically the occurrence of the MTC effect by comparing
the outflow of reaction products in the MTC system with the outflow from
a reference system with equal internal and external system parameters, but no
channel selectivity (Fig. \ref{systemPics}). The dependency of the MTC effect as a
function of the system size has been investigated in \cite{Brau03}.
The MTC effect is favored by a small number of channels and occurs only for
long channels between intersections, which by themselves lead to a very low
absolute outflow compared to a similar system with shorter channels. A recent
analytical treatment of the master equation for this stochastic many-particle
model revealed the origin of this effect at high reactivities \cite{Brza04}.
It results from an interplay of the long residence time of guest molecules
under single-file conditions with a saturation effect that leads to a depletion
of the bulk of the crystallite. Thus the MTC effect is firmly established, but the
question of its relevance for applications remains open.

Here we address this question by a systematic study of the MTC effect as a
function of the reactivity of the catalytic sites and as a function of grain
size, but using fixed small channel length. This choice is motivated by potential
relevance from an applied perspective. Moreover, for the same reason
we determine the MTC effect by making a comparison with the reference system
using the same set of fixed internal (material-dependent) parameters, but (unlike
in previous studies \cite{Karg01,Brau03,Brza04}) for each
case (MTC and REF resp.) different optimal external (operation-dependent)
parameters which one would try to implement in an industrially relevant
process. It will transpire that a significant
MTC effect (reactivity enhancement up to $\approx 30\%$) occurs in our
model system even for small channel length at realistic intermediate reaction
rates of the catalyst, provided the grain
size is sufficiently small. This may be of interest as since the first successful
synthesis of mesoporous MCM-41 nanoparticles \cite{Beck92}, there has been
intense research activity in the design and synthesis of structured mesoporous
solids with a controlled pore size. In particular, synthesis of bimodal nanostructures
with independently controlled small and large mesopore sizes has become
feasible \cite{Sun03}.\\
In this work we do not study a specific process in a specific
material, but we demonstrate the validity of the MTC concept
even if channels inside the porous material are short.
This is novel and -- from an applied viewpoint -- crucial since it is
a necessary condition for starting expensive and time-consuming
quantitative investigations in specific settings.

\section{NBK Model}

As in \cite{Brau03,Brza04} we consider the NBK lattice model \cite{Neug00}
with a quadratic array of $N\times N$ channels (Fig. \ref{systemPics}) which is
a measure of the grain size of the crystallite. Each channel has $L$ sites
between the intersection points where the irreversible catalytic process 
$A\to B$ takes place. We assume
the boundary channels of the grain to be connected to the surrounding gas
phase, modelled by reservoirs of constant densities such that the entrances of
the respective channels (extra reservoir sites) have a fixed $A$ particle 
density
$\rho$. We assume the reaction products $B$ which leave the crystallite
to be removed immediately from the gas phase such that the density of $B$
particles in the reservoir is always 0. Short-range interaction between
particles inside the narrow pores is described by an idealized hard core
repulsion which forbids double-occupancy of lattice sites.

The underlying dynamics are stochastic. We work in a continuous time description
where the transition probabilities become transition rates and no multiple
transitions occur at the same infinitesimal time unit. Each elementary transition
between microscopic configurations of the system takes place randomly with an
exponential waiting-time distribution. Diffusion is modelled by jump processes
between neighbouring lattice sites. $D$ is the elementary (attempt) rate of
hopping and is assumed to be the same for both species $A,B$ of particles.
In the absence of other particles $D$ is the
self-diffusion coefficient for the mean-square displacement along a channel. If a
neighboring site is occupied by a particle then a hopping attempt is rejected
(single-file effect). The dynamics inside a channel are thus given by
the symmetric exclusion process \cite{Spit70,Spoh83,vanB83,Schu94} which is
well-studied in the probabilistic \cite{Ligg99} and statistical mechanics
literature \cite{Schu01}. The self-diffusion along a channel is anomalous, the
effective diffusion rate between intersection points decays asymptotically
as $1/L$, see \cite{vanB83} and references therein.

At the intersections the reaction $A\to B$ occurs with a reaction rate $c$.
This reaction rate influences, but is distinct from, the effective grain
reactivity which is largely determined by the residence time of guest
molecules inside the grain which under single-file conditions grows in the
reference system with the third power of the channel length $L$
\cite{Rode99}. At the boundary sites particles jump into the reservoir with
a rate $D(1-\rho_A-\rho_B)$ in the general case. Correspondingly
particles are injected into the grain with rates $D\rho_{A,B}$ respectively.
As discussed above here we consider only $\rho_A=\rho$, $\rho_B=0$.

For the REF system A and B particles are allowed to enter and leave both types 
of channels, the vertical ($\alpha$) and horizontal ($\beta$) ones. In case of
MTC $A$($B$) particles will enter $\alpha$($\beta$)-channels only,
mimicking complete channel selectivity. Therefore all channel segments carry
only one type of particles in the MTC case. For the boundary channels
complete selectivity implies that $\alpha$-channels are effectively
described by connection with an $A$-reservoir of density $\rho_A=\rho$
($B$-particles do not block the boundary sites of $\alpha$-channels)
and $\beta$-channels  are effectively described by connection with a
$B$-reservoir of density $\rho_B=0$, respectively.
($A$-particles do not block the boundary sites of $\beta$-channels.)
This stochastic dynamics, which is a Markov process, fully defines the NBK 
model.

In both cases, MTC and REF system, the external concentration gradient 
between $A$ and $B$ reservoir densities induces a particle current inside the
grain which drives the system in a stationary nonequilibrium state.
For this reason there is no Gibbs measure and equilibrium
Monte-Carlo algorithms cannot be applied for determining steady state
properties. Instead we use dynamic Monte-Carlo simulation (DMCS)
with random sequential update. This ensures that the simulation algorithm yields 
the correct stationary distribution of the model.

\section{Monte Carlo results}

Anticipating concentration gradients between intersection points we expect due 
to the exclusion dynamics linear density profiles within the channel
segments \cite{Spoh83,Schu01,Brza04}, the slope and hence the current
being inversely proportional to the number of lattice sites $L$.
The total output current $j$ of $B$ particles, defined as the number of
$B$-particles leaving the grain per time unit in the stationary state, is the main
quantity of interest. It determines the effective reactivity of the grain.

\begin{figure}
\centerline{\includegraphics[width=7cm,angle=270]{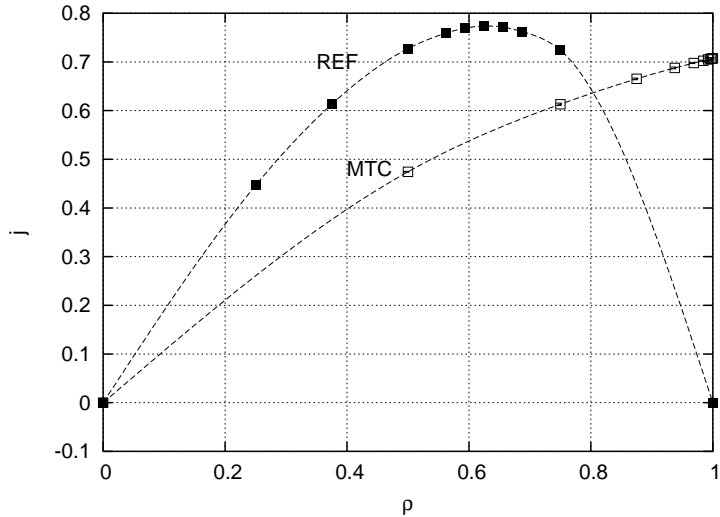}}
\caption{$j_{REF}$ (solid symbols) and $j_{MTC}$ (open symbols) as a function
of the reservoir density for a system with  $N=5$, $L=1$, $c=0.1$.}
\label{rhostar}
\end{figure}

We are particularly interested in studying the system in its maximal
current state for given reactivity $c$ and size constants $N$, $L$, which are
intrinsic material properties of a given grain. The $A$ particle reservoir
density $\rho$, determined by the density in the gas phase, can be tuned in
a possible experimental setting. Let us therefore denote the reservoir density 
which maximizes the output current by $\rho^*$ and the maximal current by $j^*$.
For MTC systems as defined above we always expect $\rho_{MTC}^*=1$,
since the highly charged entrances of $\alpha$-channels do not block the exit
of $B$-particles and hence do not prevent them from
leaving the system. Fig. \ref{rhostar} shows $j$ as a function of $\rho$ for both a 
MTC and REF system of $N=5$, $L=1$ and reactivity
$c=0.1$. Indeed for MTC the maximal output occurs for the maximal reservoir
density. In case of the REF system
$\rho_{REF}^*$ as well as $j_{REF}^*$ need to be found by simulation. We 
iteratively
approach the maximal current by a set of 9 datapoints. The "best" datapoint has 
been chosen in order to approximate the maximum. Statistical
errors are displayed. They are, however, mostly within symbol size.

In order to measure the efficiency of a MTC system over the associated REF system
we define the efficiency ratio
\begin{align}
R(c,N,L)=\frac{j_{MTC}^*}{j_{REF}^*}
\end{align}
which is a function of the system size $N$, $L$ and reactivity $c$.

\begin{figure}
\centerline{\includegraphics[width=5cm,angle=270]{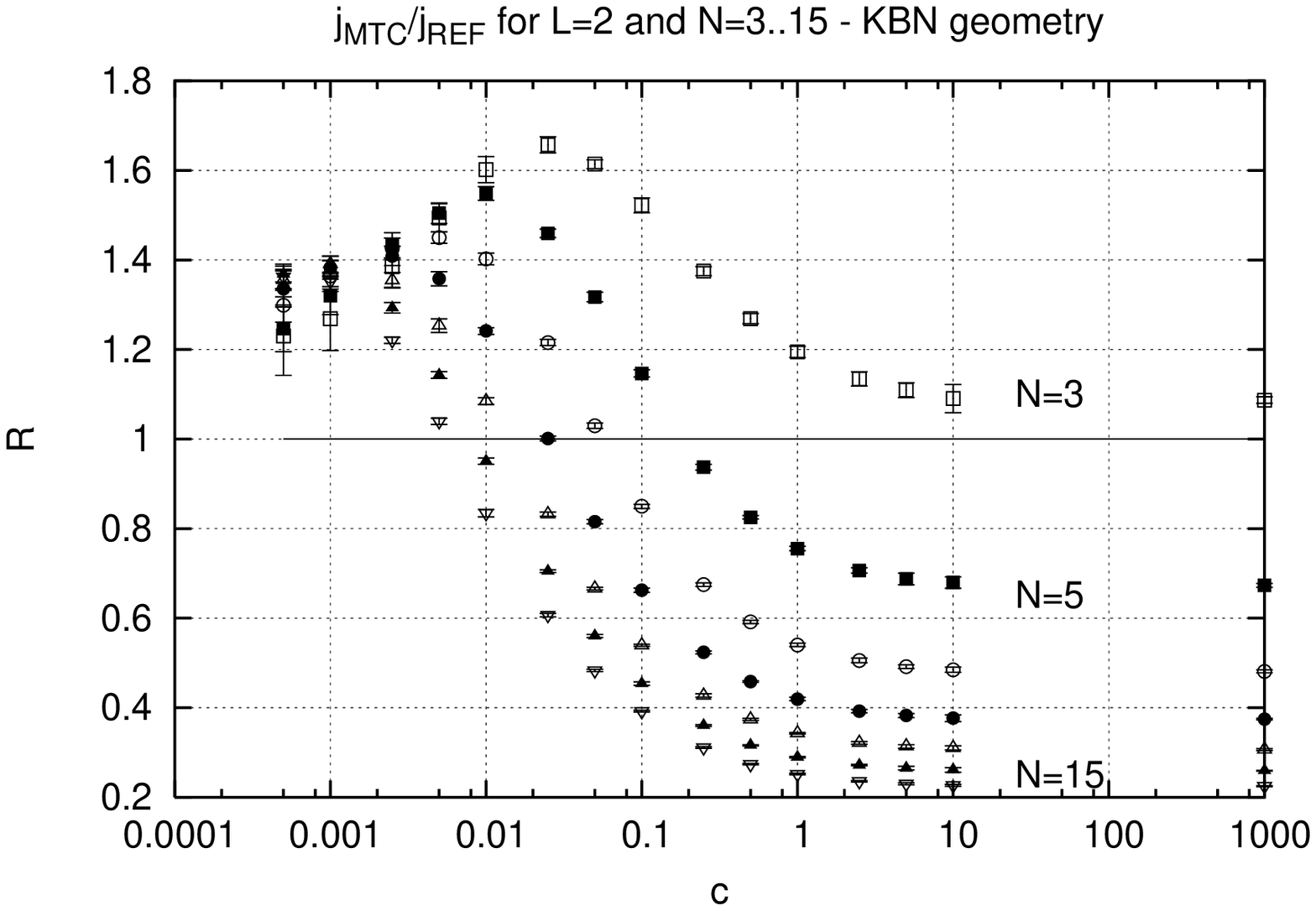}
\includegraphics[width=5cm,angle=270]{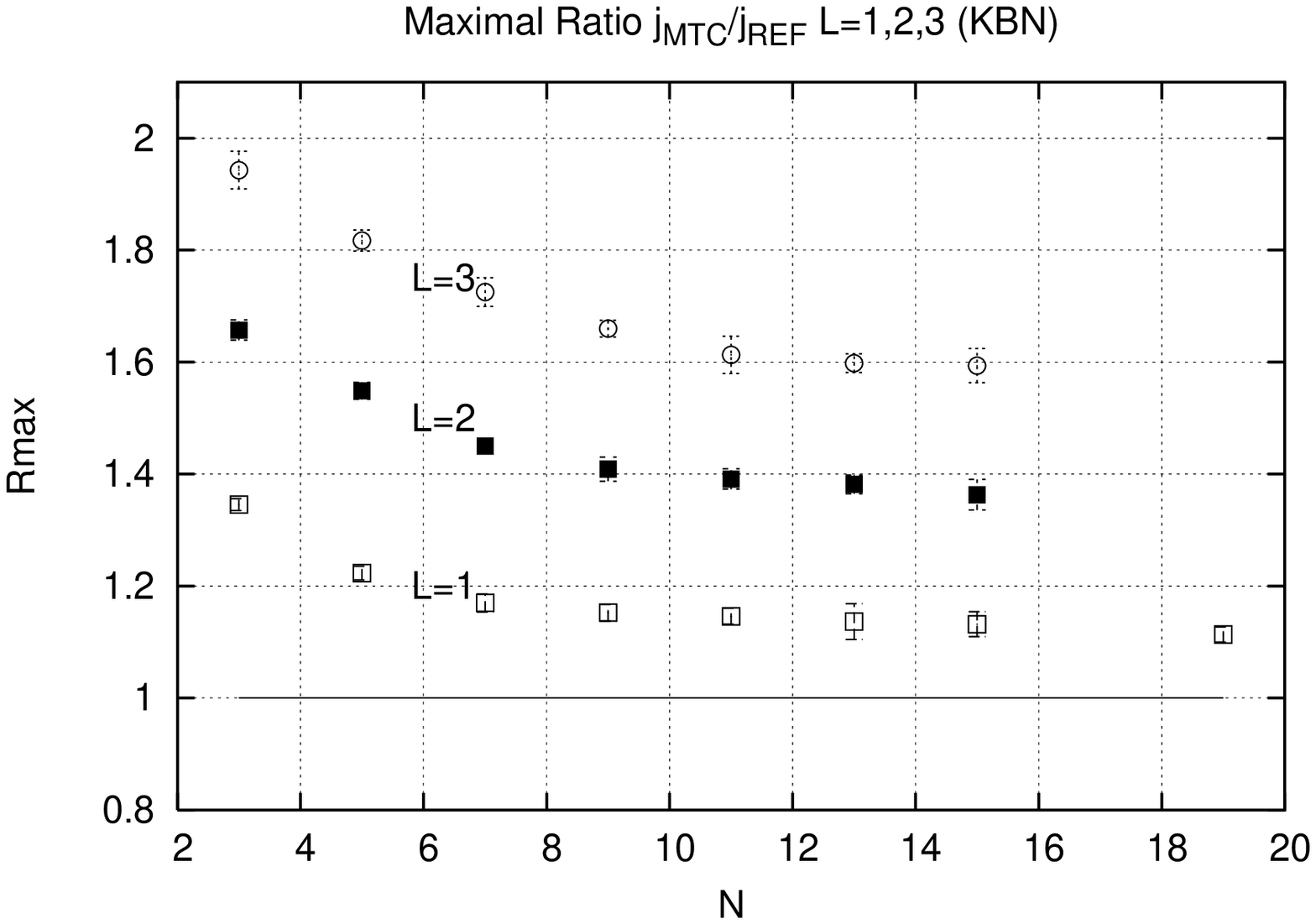}}
\caption{Left: Ratio $R(c)$ for different number of channels $N$ and $L=2$. 
Right: Maximal ratio $R^*$ for different $L$}
\label{RMTC}
\end{figure}

Fig. \ref{RMTC} (left) shows the measured ratio $R$ for a large range of 
reactivities. We plotted systems with $L=2$ and different $N$. We note that 
the MTC effect has a strong negative dependence on increasing $N$ for all 
$N$ and increasing $c$ above some optimal value $c^\ast$. We denote this
optimal value by $R_{max}$. Fig. \ref{RMTC} (right)  shows $R_{max}^*(N)$ for 
different $L$ and proves that there is an MTC effect  for any $L$ and any $N$. 
Notice, however, that with increasing $N$ the optimal ratio not only decreases,
but appears at unnaturally small reactivities $c$. This is highly undesirable as 
then the actual output current shrinks to zero. Even though more efficient than 
a REF system, a MTC grain with large $N$ would not operate under practically
relevant conditions.

\begin{figure}
\centerline{\includegraphics[width=7cm,angle=270]{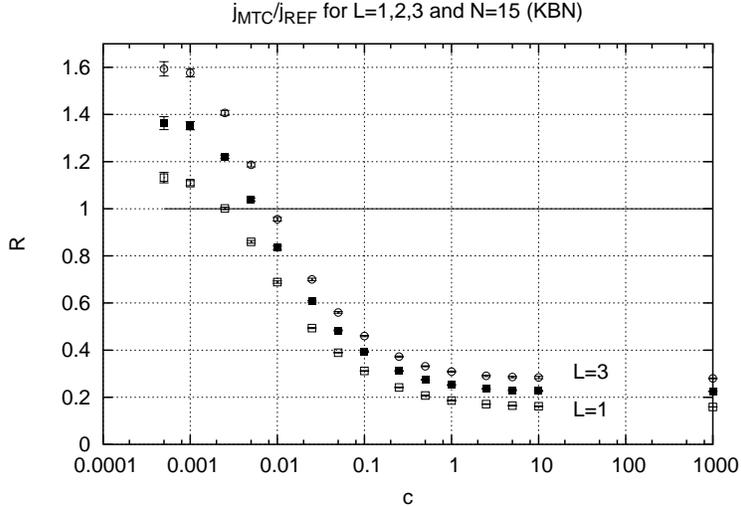}}
\caption{Dependence of the output ratio on the channel length $L$.}
\label{increasingL}
\end{figure}

As expected from previous results \cite{Brau03,Brza04}  the MTC effect
is seen to increase with increasing $L$, even though the measure $R$ used here is
different (Fig. \ref{increasingL}). This follows from theoretical studies of
single-file systems. The mean traveling time of a product molecule
through a channel of length $L$ is proportional
to $L^2$ in the MTC case as in ordinary diffusion, but proportional to $L^3$
in the REF case due to mutual blockage. Hence the current is proportional to
$1/L$ in a MTC system, but proportional to $1/L^2$ in a REF system. This
holds for all values of the parameters and hence for sufficiently large $L$
the MTC system becomes more efficient.

\begin{figure}
\centerline{
\includegraphics[width=5cm,angle=270]{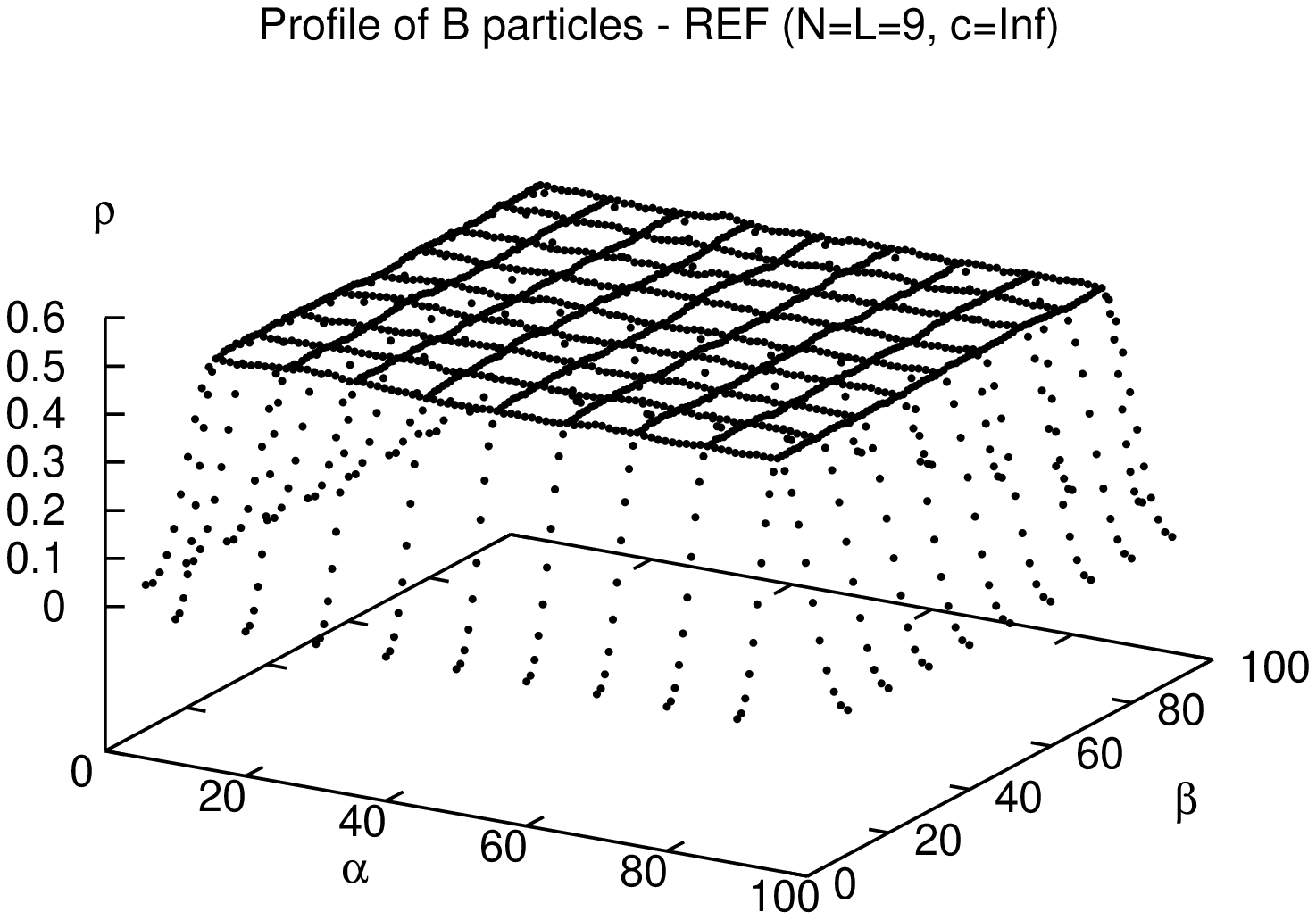}
\includegraphics[width=5cm,angle=270]{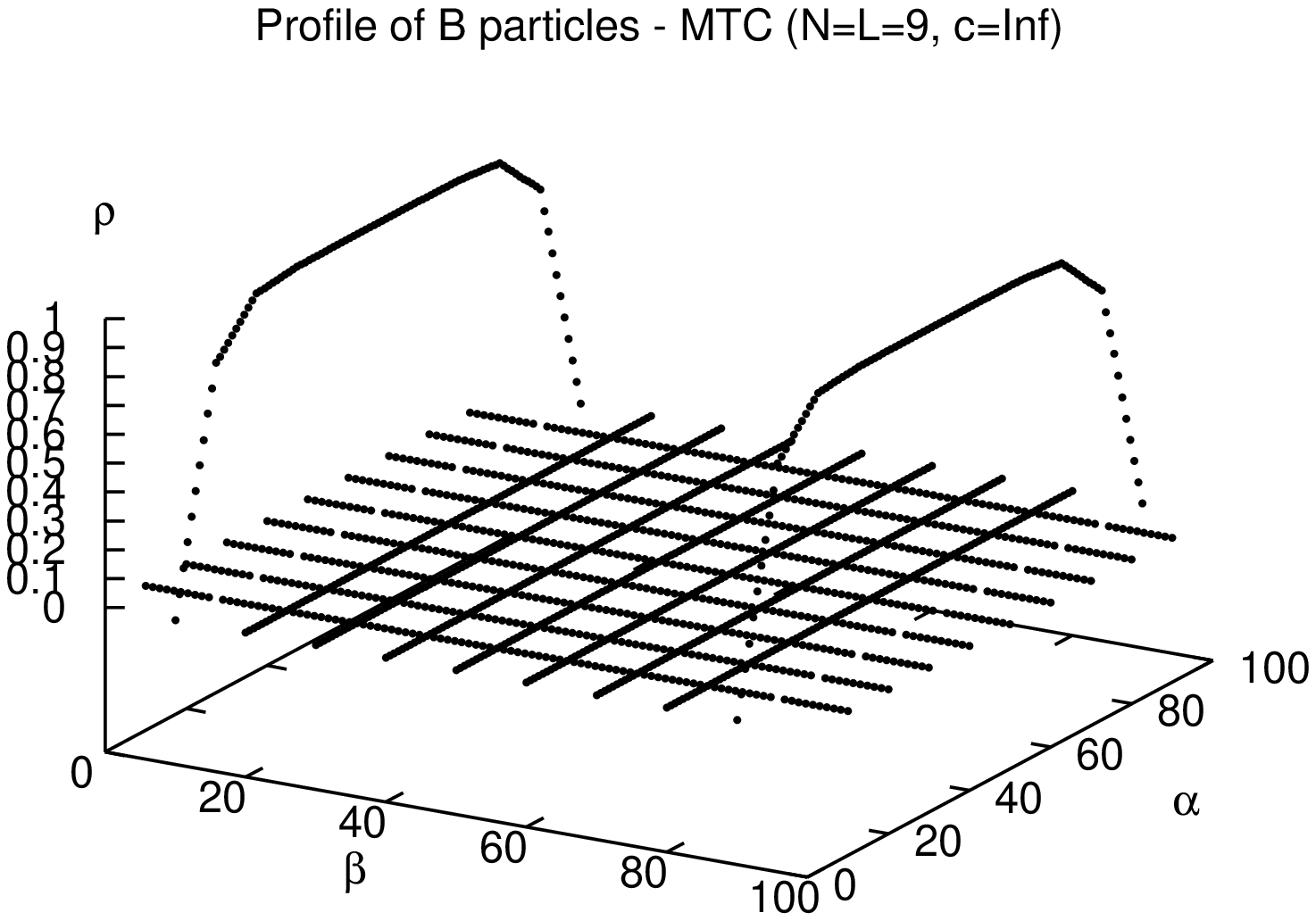}
}
\caption{Profiles of the REF (left) and MTC (right) system in the 
large-reactivity case. $N=L=9$}
\label{profiles}
\end{figure}

\begin{figure}
\centerline{\includegraphics[width=6cm,angle=270]{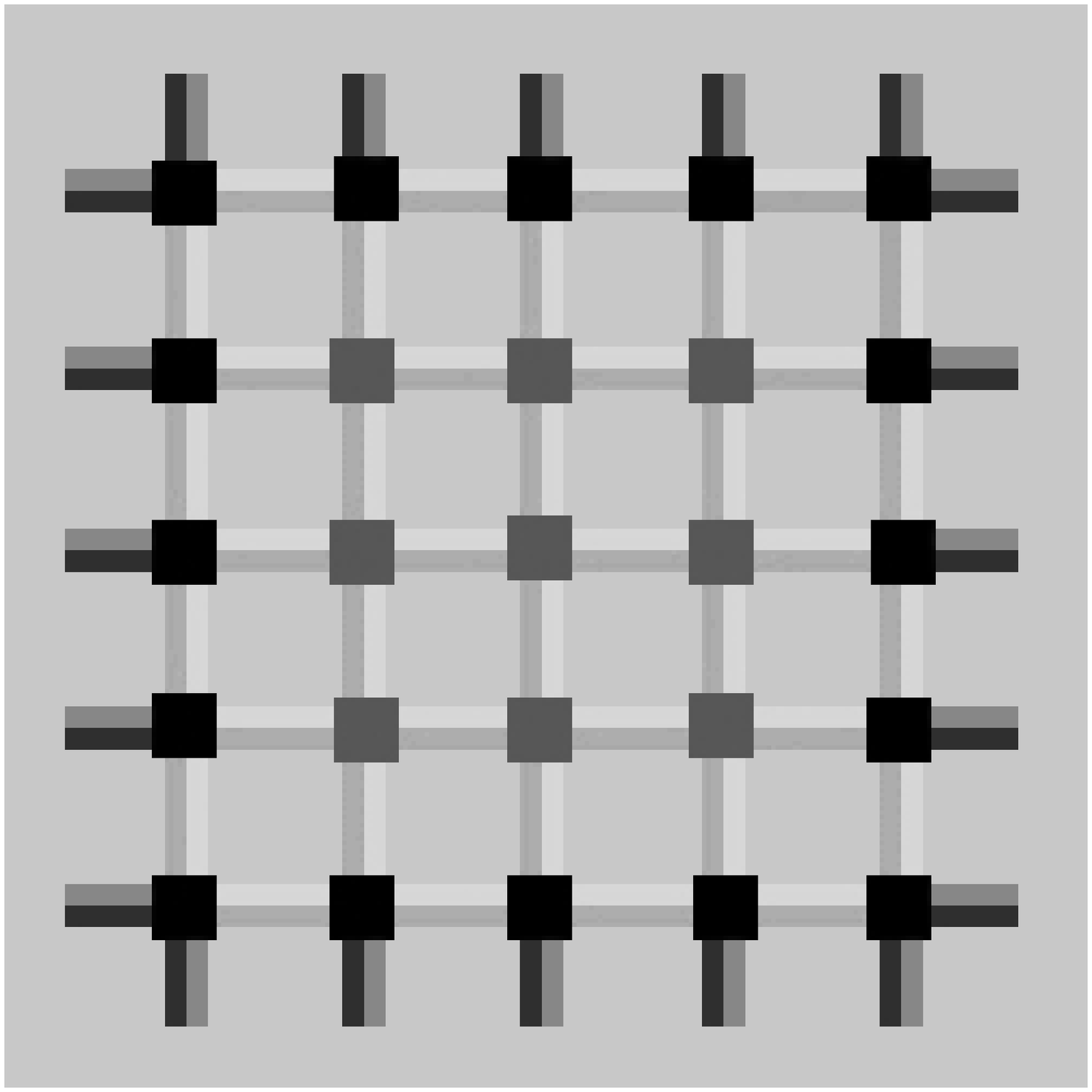}
\includegraphics[width=6cm,angle=270]{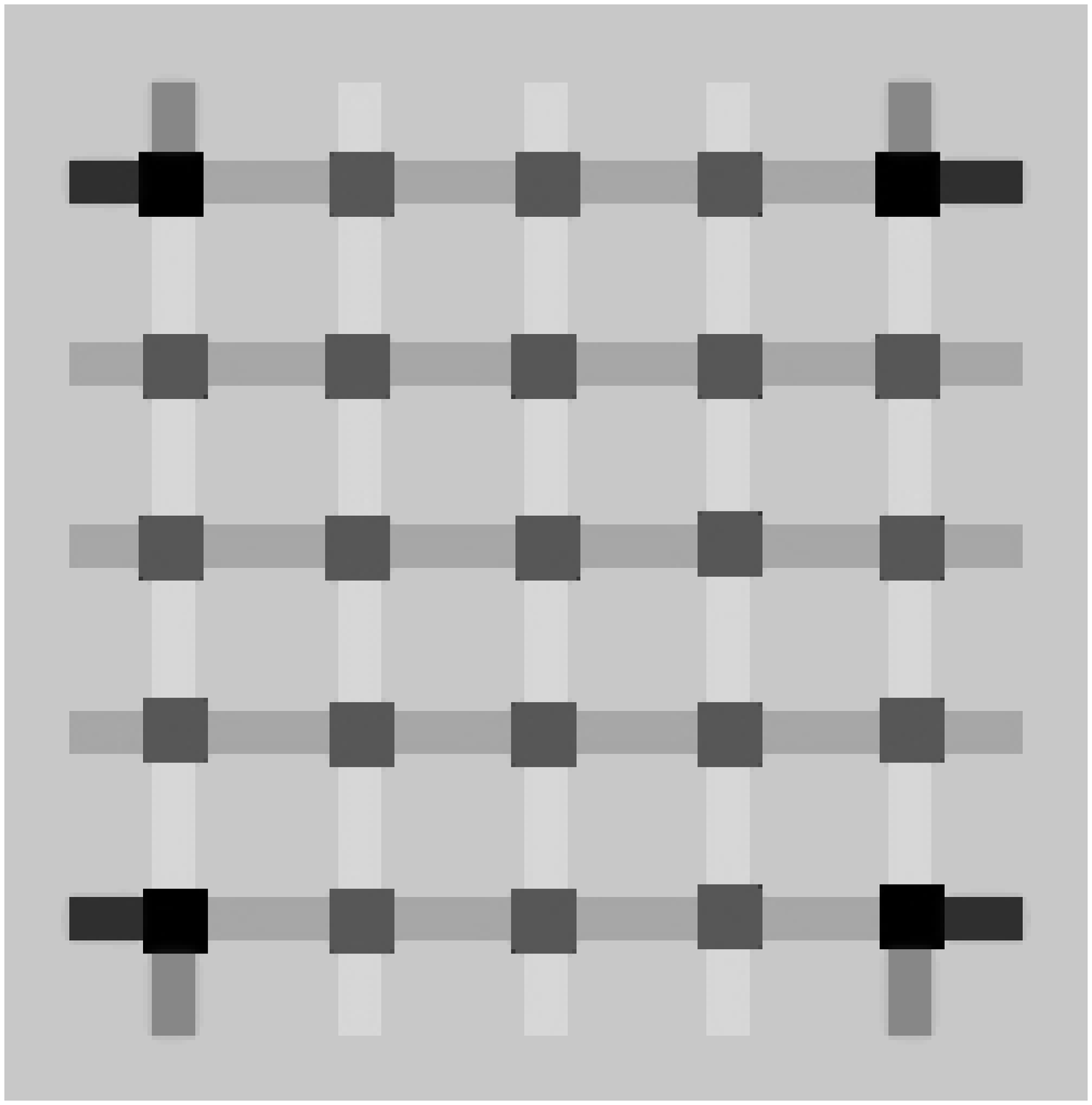}}
\caption{Active channel segments in the large-reactivity case. 
REF system (left), MTC system (right)}
\label{max}
\end{figure}

In order to understand the strong decrease of $R$ for large reactivities and
large number of channels it is instructive to study the stationary density
profiles. Fig. \ref{profiles} shows the $B$ particle densities for a REF system
(left) and MTC system (right) with $N=L=9$ and large reactivity $c\to\infty$.
Theoretical investigations of MTC systems in the large reactivity case
\cite{Brza04} show that in the state of maximal current ($\rho=1$) the output of
$B$ particles is independent of the number of channels. For large $L$ the
maximal output current becomes $j^*_{c\to\infty} \simeq \frac{4D}{L}$ where
$D$ is the diffusion constant. A nonvanishing current of $B$ particles can be
observed only at the four corners of the lattice (Fig. \ref{max} right).

For fixed moderate $c$ this extreme situation is not realized, but nevertheless
with increasing $N$ one expects that the bulk gets increasingly depleted,
since in each layer a fraction of $A$ particles is converted into $B$
particles. Thus the total $A$-density in each layer may be described by the form
\begin{align}
\frac{d}{dx} N_A(x) = - \gamma N_A(x)
\end{align}
The coarse-grained ansatz for the number $N_A(x)$ of $A$-particles in layer $x$ predicts an
exponential decrease of the $A$ density,
leaving only an active boundary layer of finite thickness
\begin{align}
\xi = 1/\gamma \propto 1/c
\end{align}
at the top and
bottom respectively of the (in our simulation two-dimensional) grain.
Hence, as a function of $N$, $j^\ast_{MTC}$ saturates at some constant
\begin{align}
\lim_{N \to\infty} j^\ast_{MTC}(c,N,L)= C^\ast_{MTC}(c,L).
\end{align}

On the other hand, in the REF system the output current scales linearly with
increasing $N$ for all, even large, $c$ (Fig. \ref{max} left). This is because
even though again the bulk depletes with increasing $N$ the active boundary
layer is a surface scaling linearly with $N$. Thus
\begin{align}
\lim_{N \to\infty} j^\ast_{REF}(c,N,L)= N C^\ast_{REF}(c,L)
\end{align}
Hence
\begin{align}
R(c,N,L) \propto 1/N
\end{align}
and the MTC effect vanishes at some $N$ for fixed reactivity $c$
and channel length $L$.

\begin{figure}
\centerline{
\includegraphics[width=5cm,angle=270]{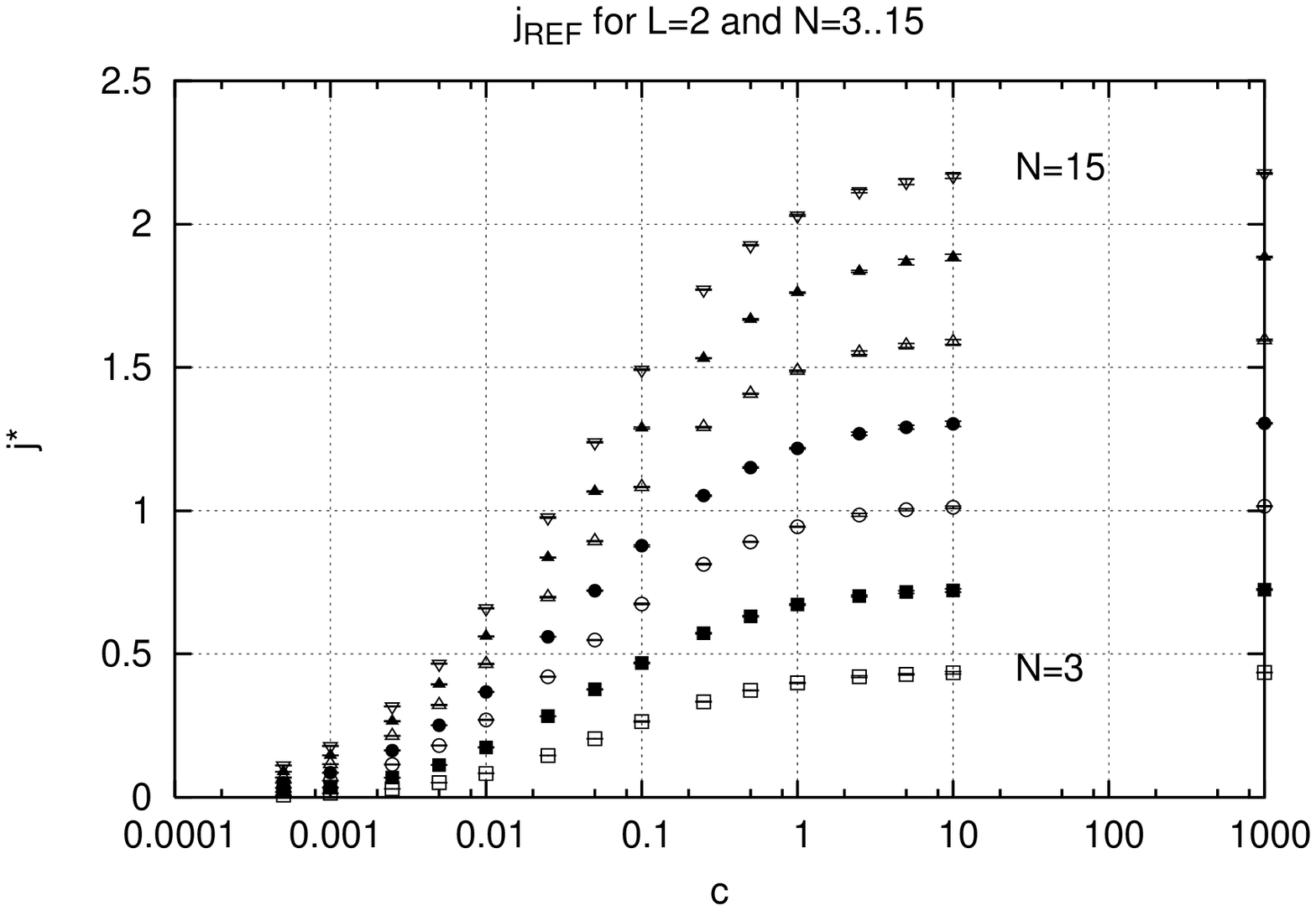}
\includegraphics[width=5cm,angle=270]{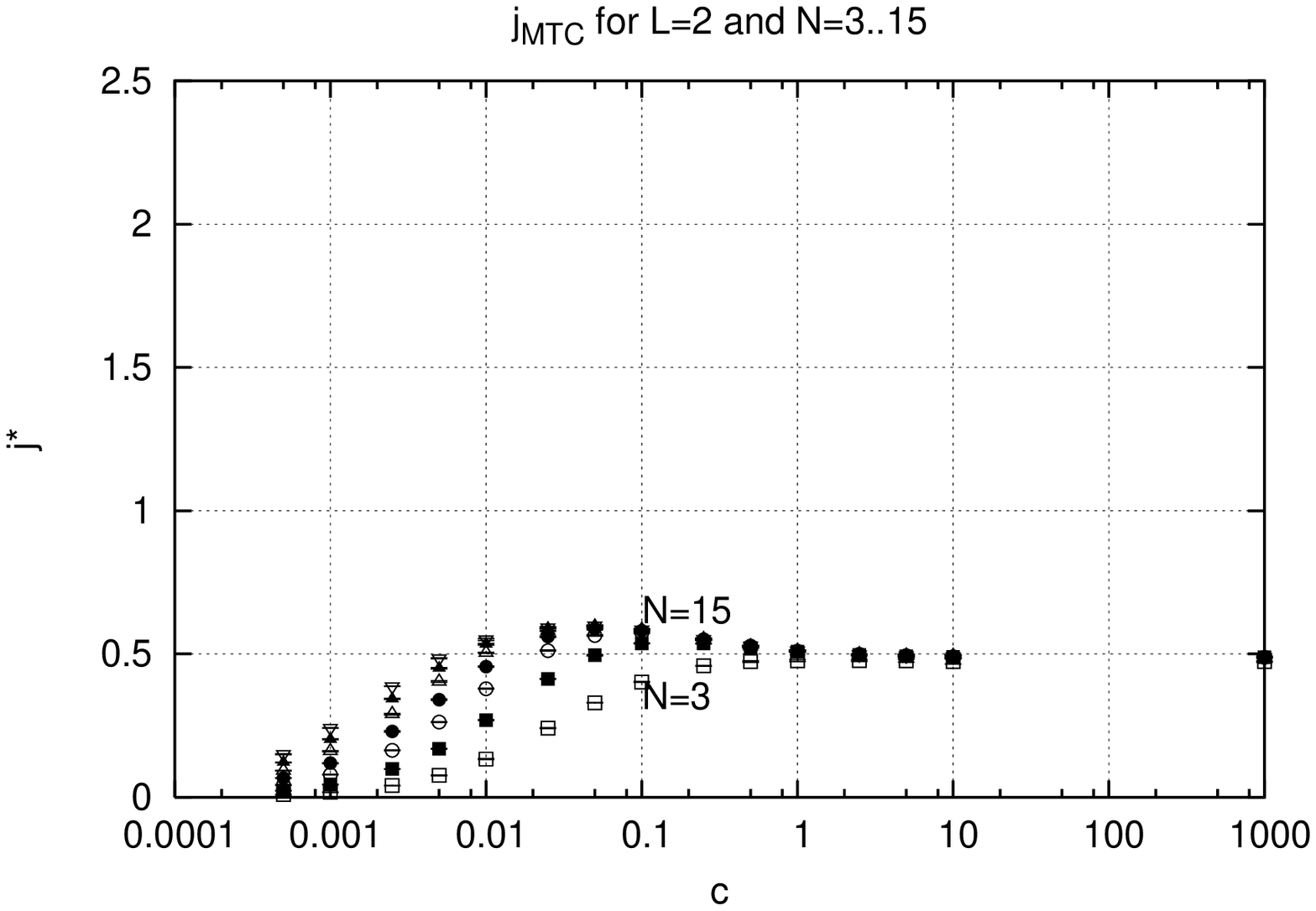}
}
\caption{Maximal currents for the REF system (left) and the MTC system 
(right). $L=2$ and different $N$.}
\label{currents}
\end{figure}

Looking on a wider range of reactivities (Fig. \ref{currents}) we notice that
the maximal currents for MTC systems saturate for some intermediate $c$
whereas in the REF case the current reaches a plateau for large reactivities.
This observation can be rationalized by noticing that an increase of the
output with increasing $c$ is limited by the incoming current of available
$A$ particles. Since in the REF system $A$ and $B$ particles block each
other an increasing current of $B$-particles always restricts the number of
$A$ particles diffusing in. Hence the saturation due to high reactivity
sets in only for larger values of $c$ than in the MTC system.


\section{REF with small reactivities}

The arguments put forward above for explaining qualitative and quantitative
features of the MTC effect at intermediate and large reactivities $c$ fail for
small reactivities, i.e., when $c$ is of the order of the inverse mean
intracrystalline residence time. We first consider the reference system.

In this case the grain contains only a very small number of $B$ particles at
any time.
In this low-concentration regime the diffusion of $B$ particles
inside the grain may be
described by a linear diffusion equation with an effective diffusion coefficient
determined by the interaction with $A$ particles as follows. The 
$A$ particles are considered as a medium with constant equilibrium density 
$\rho$ throughout the system. At the intersections $B$ particles are created 
randomly with effective rate $c\rho$. We then describe the
diffusion of a
$B$ particle in the channel between intersections by a random walk
from intersection to intersection with an effective diffusion constant $D_{eff}$
given by the (inverse) mean travelling time between intersections. 
Let $\rho_{(x,y)}$ be the $B$ particle density at the intersection denoted by
$(x,y)$ and $\Delta$
a discrete two-dimensional Laplace operator
\begin{align}
\label{laplace}
\Delta\rho_{(x,y)}=\frac{1}{4}\left( \rho_{(x+1,y)}+\rho_{(x-1,y)}+
\rho_{(x,y+1)}+\rho_{(x,y-1)}-4\rho_{(x,y)}\right).
\end{align}
Here the lattice unit is given by the channel length rather than the
pore size inside a channel.
For the $B$ particle density at intersections we thus obtain
\begin{align}
\label{diffeq}\frac{\partial}{\partial t}\rho_{(x,y)}=D_{eff}\Delta\rho_{(x,y)}+c\rho.
\end{align}
A stationary solution to \eqref{diffeq} can be found by use of the
discrete sine transform
$\tilde{\rho}_{(q,p)}=\sum_{x=1}^N\sum_{y=1}^N \rho_{(x,y)}
\sin\frac{q\pi x}{N+1}\sin\frac{p\pi y}{N+1}$.
We express \eqref{diffeq} (with vanishing time derivative) in terms of 
the  transformed density
$\tilde{\rho}_{(q,p)}$. Taking into account the boundary conditions
$\rho_{0,y}=\rho_{x,0}=\rho_{x,N+1}=\rho_{N+1,y}=0$ with 
$0\le x,y \le N+1$ we find
\begin{align}
\label{diffeqtrns}
\tilde{\rho}_{(q,p)}=\frac{2c\rho_A}
{\cos\frac{q\pi}{N+1}+\cos\frac{p\pi}{N+1}-2}
\sum_{n=1}^N\sum_{m=1}^N\sin\frac{q\pi n}{N+1}\sin\frac{p\pi m}{N+1}.
\end{align}
The non zero contributions of the double sum can be expressed as a product 
of two Cotangents. Transforming back finally yields
\begin{align}
\label{REFsolution}
{\rho}_{(x,y)}&=\lambda\sum_{n=1}^N\sum_{m=1}^N
\frac{B_{(n,m)}}{\cos\frac{n\pi}{N+1}+\cos\frac{m\pi}{N+1}-2}
\sin\frac{n\pi x}{N+1}sin\frac{n\pi y}{N+1}\\
B_{(n,m)}&=
\begin{cases}
0 & \text{if $n$ or $m$ even} \\
\frac{1}{(N+1)^2}\cot\frac{m\pi}{2(N+1)}\cot\frac{n\pi}{2(N+1)} & \text{else}
\end{cases}
\end{align}

$D_{eff}$ together with the reactivity and reservoir density, is absorbed 
into a fitting parameter
$\lambda \sim \frac{c\rho}{D_{eff}}$. With ${\rho}^{Sim}_{(x,y)}$ being the 
$B$ particle densities obtained
from simulations and ${\rho}^{Th}_{(x,y)}$ the theoretical densities, we define
the homogenized mean square deviation
\begin{align}
\label{quality}
Q:=\frac{2}{N}\sqrt{\sum_\text{intersections} 
\left(\frac{{\rho}^{Sim}_{(x,y)}-{\rho}^{Th}_{(x,y)}}{{\rho}^{Sim}_{(x,y)}+{\rho}^{Th}_{(x,y)}}\right)^2}
\end{align}
as a measure for the quality of the approximation. The sum runs over all 
intersections, with the local deviation normalized by the local mean.
This assures that all intersections contribute with their proper weight.

\begin{figure}
\centerline{\includegraphics[width=5cm,angle=270]{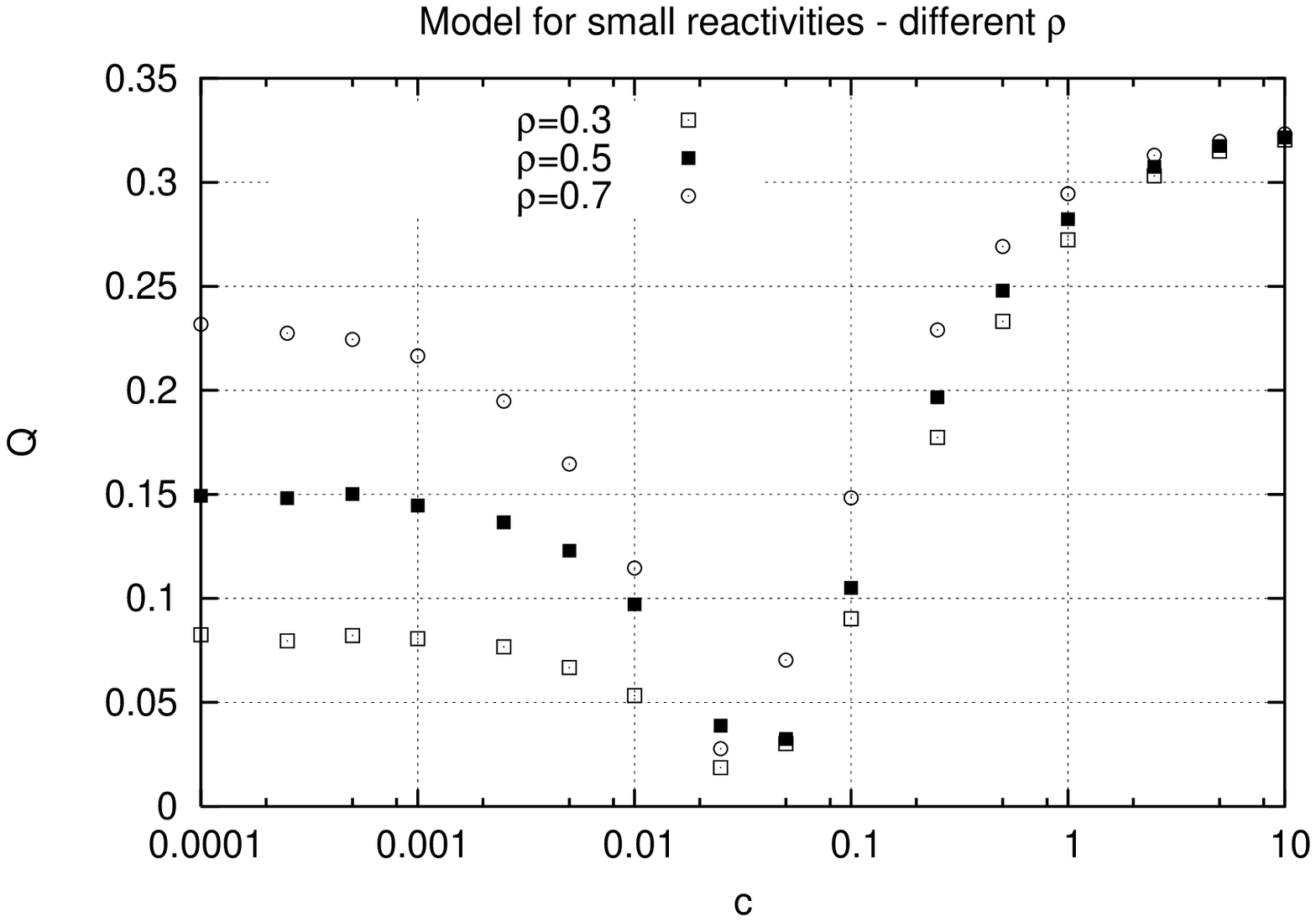}
\includegraphics[width=5cm,angle=270]{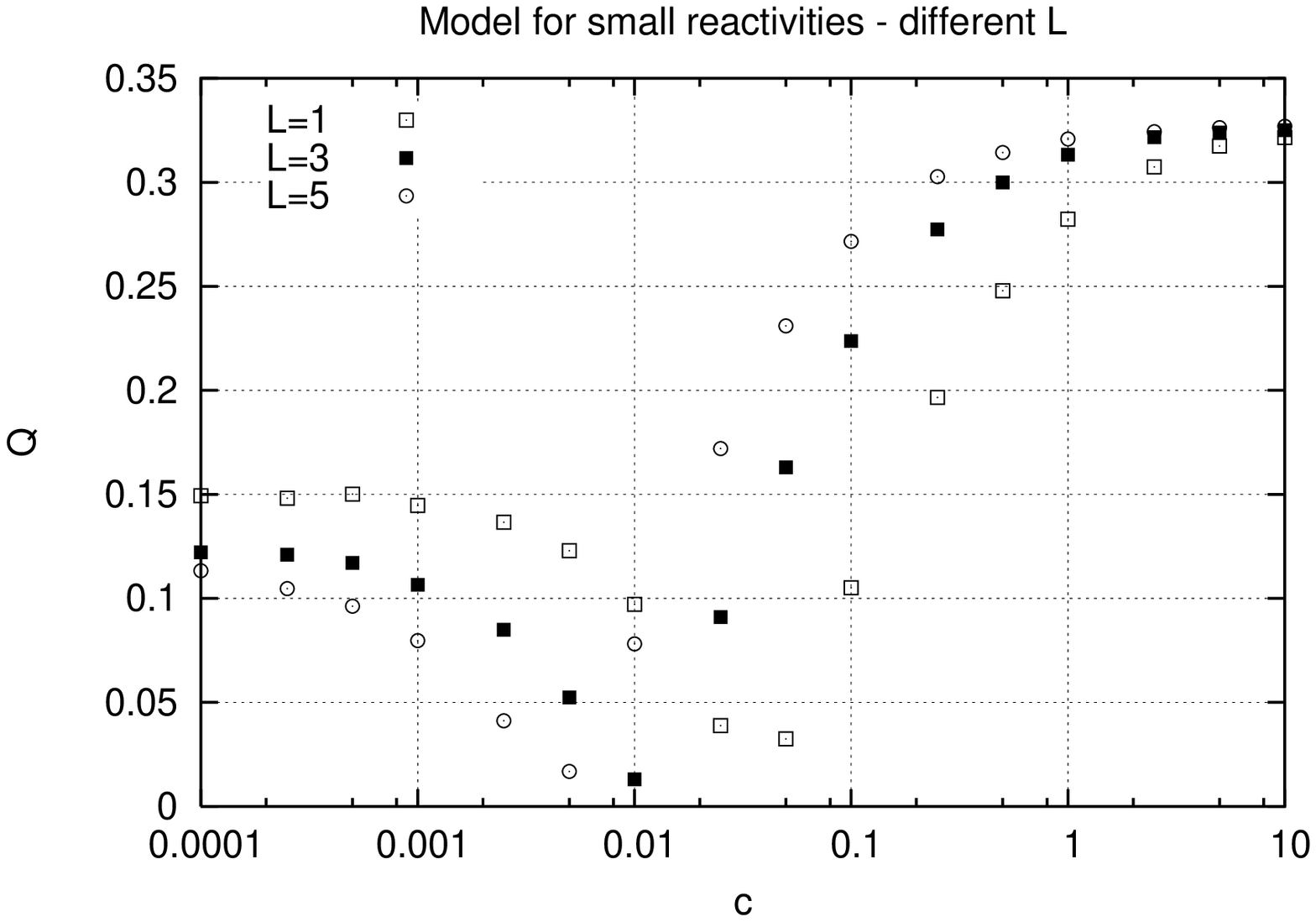}}
\caption{Collapse of the simulated and calculated profiles for REF systems. 
Left: $N=5$, $\rho=0.5$ and different $L$. (right) $N=5$, $L=1$ and different 
$\rho$.}
\label{csmall1}
\end{figure}

For large $c$ the profile flattens (see Fig. \ref{profiles}) and is not
very well described by \eqref{REFsolution}. Fig. \ref{csmall1} shows $Q$ 
as a function of $c$. The boundary density has been chosen to ${\rho}=0.5$. 
For small reactivies the collapse of the simulated and calculated profile
is fairly good ($Q\approx 0.1$). However, the best collapse occurs for 
small intermediate reactivities. This somewhat peculiar
behavior can be explained by assuming that this random walk is not a Markov 
process as implied in the derivation of \eqref{REFsolution}. The structural 
change in the occupation of the channel segment, once a particle covered the 
distance between two adjacent intersections, implies a "memory" effect. 
Hence we cannot assume perfect statistical independence of subsequent random 
traveling times between intersections which implies that the Markov assumption
is not very well satisfied. As we increase the reactivity
more than one B particle may be present in the system. This corresponds to 
an ensemble average over almost independent random walkers which 
cancels the time correlations of each individual
walker and thus improves the validity of the Markov assumption. This leads to
the good data collapse for some intermediate reactivities. As $c$ increases 
further both the assumption of equilibrium of $A$ particles and the 
low-concentration approximation (\ref{diffeq}) for $B$ particles fail.\\

\section{MTC with small reactivities}

\begin{figure}
\centerline{
\includegraphics[width=5cm,angle=270]{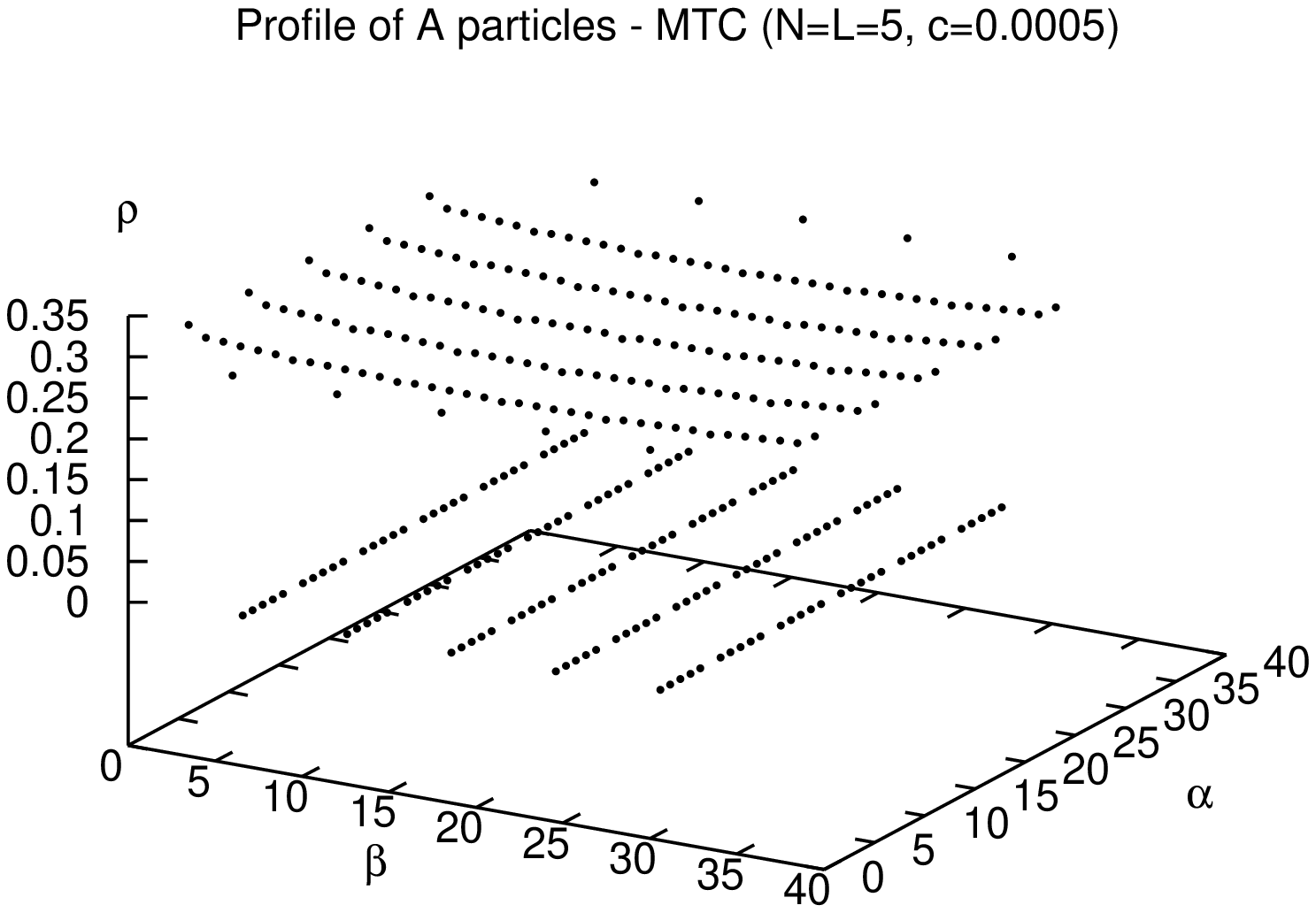}
\includegraphics[width=5cm,angle=270]{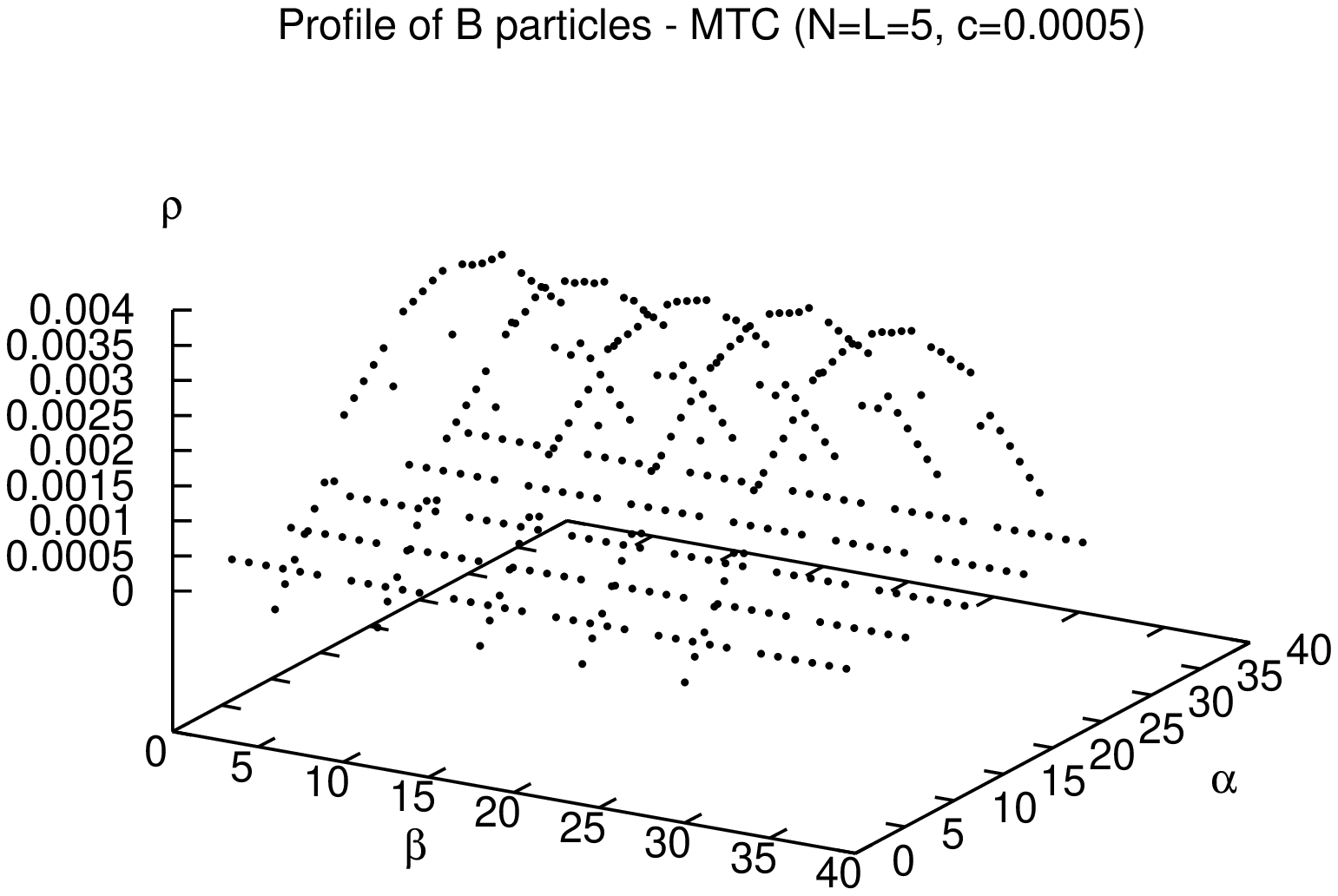}
}
\caption{Profiles for a MTC system $N=L=5$ and small reactivity $c=0.005$. 
(left) densities of A and (right) of B particles. $\rho=0.3$}
\label{MTCprofile}
\end{figure}

Fig. \ref{MTCprofile} shows the stationary density profiles of a MTC system
with small reactivity. We first note that the output current is proportional to the
number $N$ of $\beta$ exit channels, in agreement with the observation
that $R_{max}$ approaches a constant for large $N$, rather than decaying
proportional to $1/N$ as $R$ for fixed reactivity does. Moreover, due to
rare transition events all $\alpha$-channels are almost in equilibrium
with the reservoir (Fig. \ref{MTCprofile} left). Thus it is sufficient to single out only 
one $\beta$ channel.

\begin{figure}
\centerline{
\includegraphics[width=12cm]{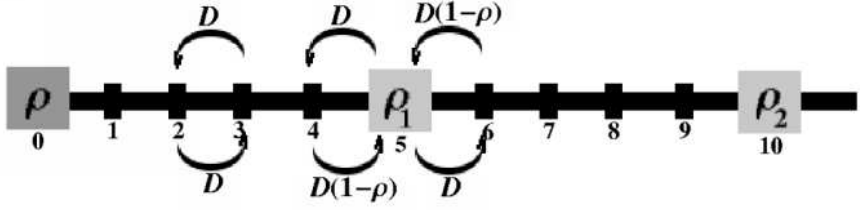}
}
\caption{$\beta$ channel with hopping rates. $L=4$}
\label{MTCbeta}
\end{figure}

We adapt the approach we took for the REF system to the present
case. Due to low concentration we use a linear
diffusion equation for the $B$-density inside a $\beta$-channel. 
There are, however, two essential differences.
(1) Because auf the absence of $A$-particles we do not consider the
occupation of intersections alone, but we track the motion of $B$
particles inside a channel. (2)
Due to the inequivalence of the sites inside the $\beta$-channels and the
intersection points where also $A$ may be located,
we need to describe a random walker with space dependent hopping rates. 
Intersections, on the one hand, serve as sites of B particle creation with a 
rate proportional to the reactivity. On the other hand intersections are occupied 
by A particles with a probability $\rho$ and hence, block B particles. This 
leads to different hopping rates onto and from an intersection as
displayed in Fig. \ref{MTCbeta}.

The master equation description leads to a set of equations for the
local $B$ particle densities $\lav n_x \rav$.
\begin{align}
\label{densities}
\frac{d}{dt}\lav n_x \rav&=
\begin{cases}
D\left( \lav n_{x-1} \rav + \lav n_{x+1} \rav - \lav n_{x} \rav - (1-\rho)\lav n_{x} \rav \right) &x=(L+1)r\pm1\\
D\left( 1-\rho \right)\left( \lav n_{x-1} \rav + \lav n_{x+1} \rav \right)-2D\lav n_{x} \rav + c\rho &x=(L+1)r\\
D\left( \lav n_{x-1} \rav + \lav n_{x+1} \rav - 2\lav n_{x} \rav \right) &\text{else}\\
\end{cases}
\end{align}
Here $x$ is the lattice position inside a channel in units of the pore size.
In the stationary state left hand site of \eqref{densities} vanishes
and it follows a recurrence relation for the B-particle intersection
densities $\rho_r\equiv\lav n_{(L+1)r}\rav$.
\begin{align}
\label{MTCrecurence}
\rho_{r}&=\frac{1}{2}\left(\rho_{r+1}+\rho_{r-1}+K \right)\\
K&=\frac{c\rho}{D}\left[ \left(L+1\right) - \left( L-1 \right)\rho \right]
\end{align}

\begin{figure}
\centerline{\includegraphics[width=8cm,angle=270]{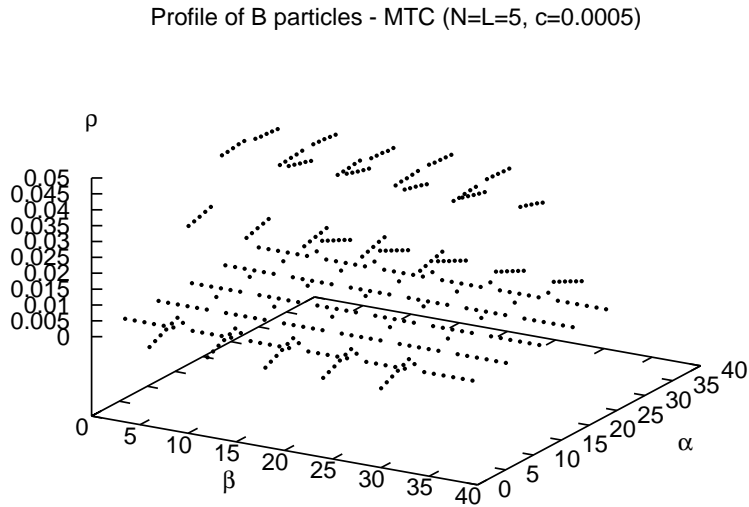}}
\caption{Demonstration of "jumps" at intersections for large $\rho$. Here we 
choose $\rho=0.9$.}
\label{MTCsmalldemo}
\end{figure}

The inhomogeneity $K$ depends on the segment length $L$ and the transition
rates. A solution satisfying \eqref{MTCrecurence} and the boundary
condition $\rho_0=\rho_{N+1}=0$ can be obtained by use of
a quadratic ansatz. We find
\begin{align}
\label{MTCsolution}
\rho_r\equiv\lav n_{(L+1)r} \rav=\frac{1}{2}rK\left[N+1-r\right]
\end{align}
Due to the modified hopping rates in the neighborhood of the intersections the 
density profile is linear between $\lav n_{(L+1)r-1}\rav$ and 
$\lav n_{(L+1)r-L}\rav$ rather than between
the intersections $\rho_r$ and $\rho_{r-1}$ itself. This becomes noticeable for
large reservoir densities $\rho$. There is a "jump" between intersections and 
their
adjacent lattice sites. Fig. \ref{MTCsmalldemo} illustrates this by means of a
profile obtained from MC simulations). Let us consider a channel segment 
embedded by the two intersections $\rho_r$ and $\rho_{r-1}$. Solving 
\eqref{densities} for the lattice sites located next to the intersections leads 
to
\begin{align}
\label{MTCnextsites}
\lav n_{(L+1)r-1} \rav &=\frac{\rho_{r-1}+\rho_r\left( L-(L-1)\rho \right)}
                                         {(L+1)-2L\rho+(L-1)\rho^2}\\
\lav n_{(L+1)r-L} \rav &=\frac{\rho_{r}+\rho_{r-1}\left( L-(L-1)\rho \right)}
                                         {(L+1)-2L\rho+(L-1)\rho^2}
\end{align}
and for the very first channel segment we find
\begin{align}
\label{MTCboundsegment}
\lav n_{L} \rav &= \frac{L\rho_{1}}{L+1-L\rho}.
\end{align}
We have now fully determined the profile for MTC systems with small reactivity.
The currents
between neighboring lattice sites are proportional to the density difference
$j_{(x,x+1)}\sim (\lav n_{x+1} \rav - \lav n_x \rav )$ with a proportional 
constant being the actual hopping rate.

\begin{figure}
\centerline{
\includegraphics[width=8cm,angle=270]{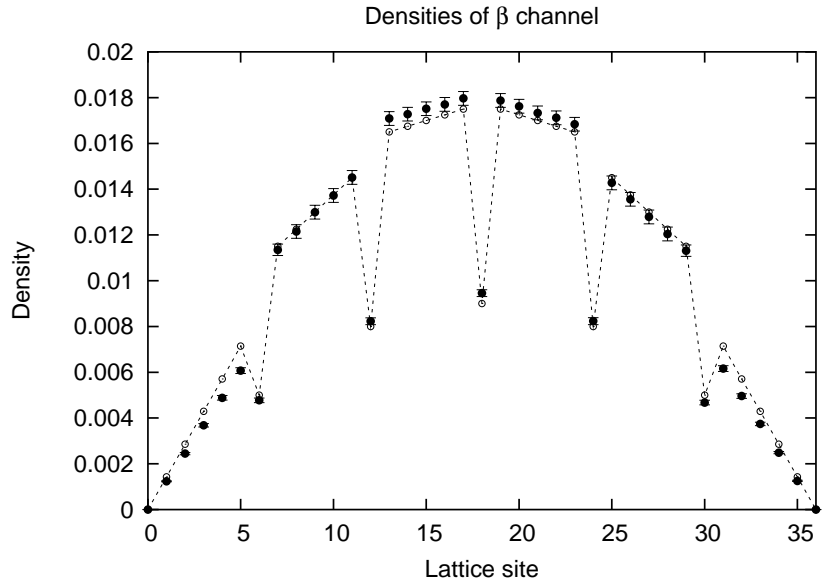}}
\caption{Densities of a $\beta$ channel. Small open circles connected with the
dotted line show the theoretical densities.
Solid circles are densities obtained by MC simulation 
($N=L=5$, $c=0.0005$, $\rho=0.5$) .}
\label{MTCprofiledemo}
\end{figure}

\begin{figure}
\centerline{
\includegraphics[width=6cm,angle=270]{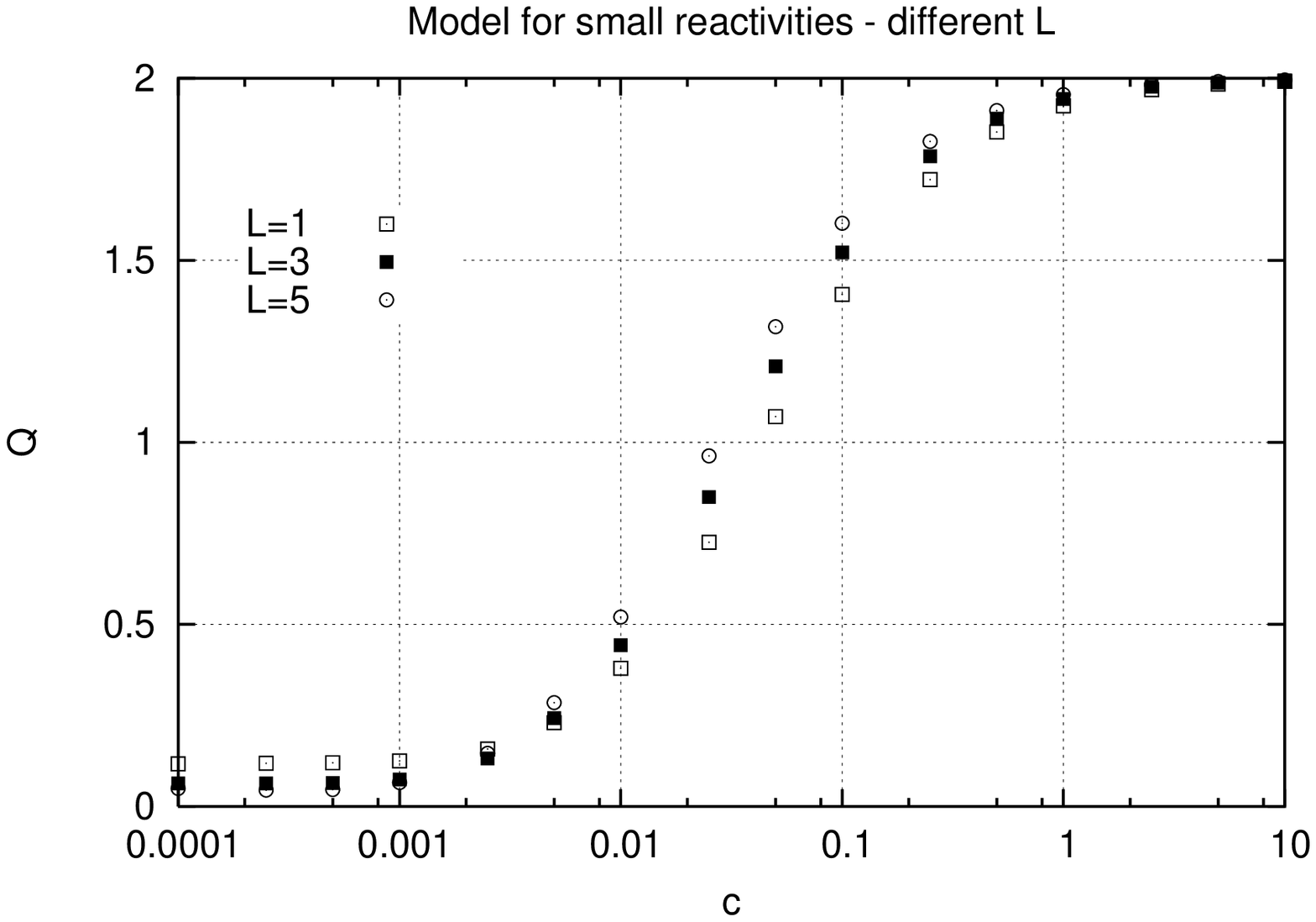}
\includegraphics[width=6cm,angle=270]{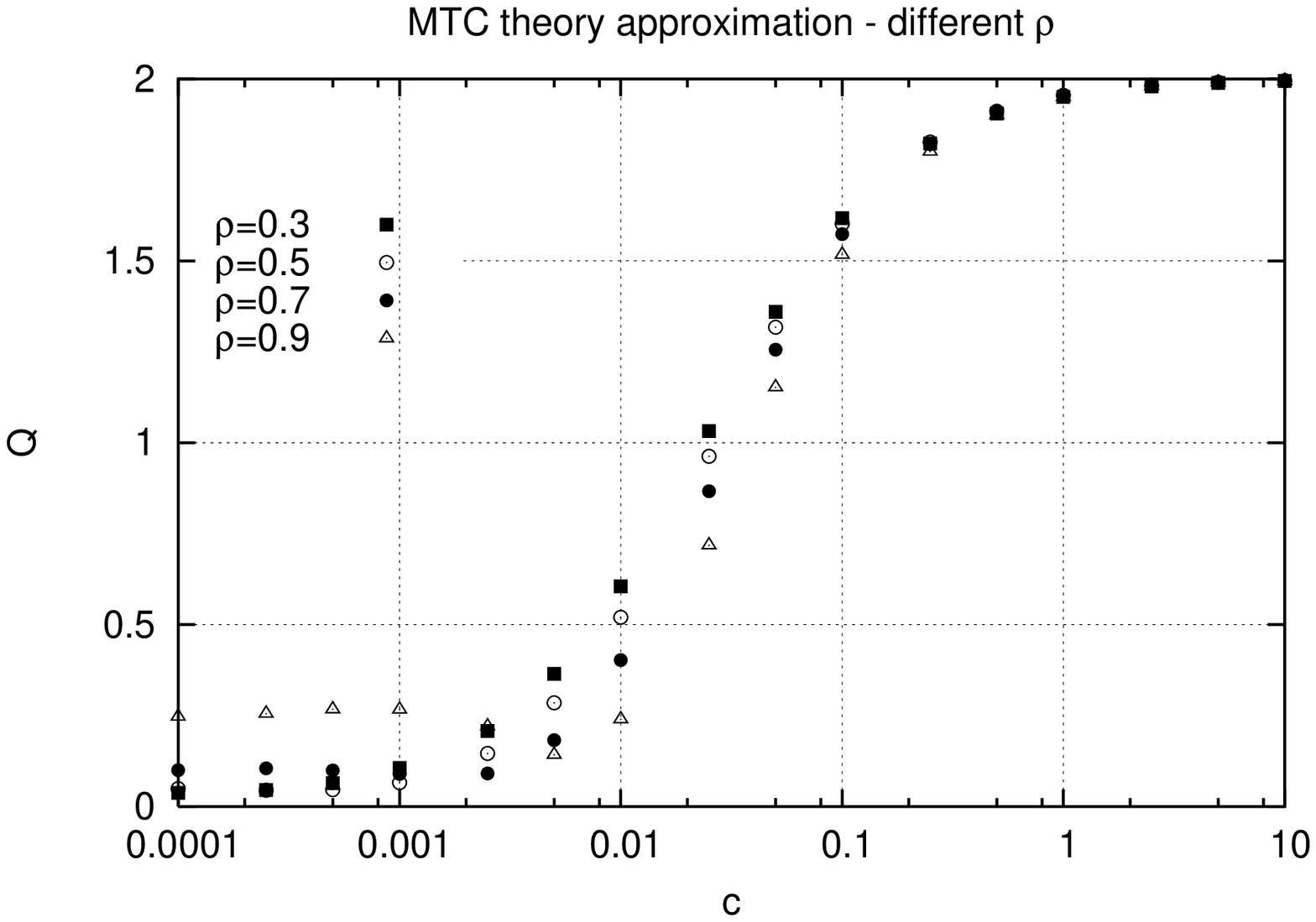}
}
\caption{Collapse of the simulated and calculated profiles for MTC
systems. Left: $N=5$, $\rho=0.5$ and
different $L$. Right: $N=5$, $L=5$ and different $\rho$.}
\label{MTCparameters}
\end{figure}

Fig. \ref{MTCprofiledemo} shows the simulated profile of a $\beta$ channel 
together with the theoretical densities. We chose an intermediate reservoir 
density $\rho=0.5$. The collapse of the two curves
is very good. In order to study the quality of the approximation for different 
sets of parameter we use the definition \eqref{quality} above. The normalized 
mean square deviation $Q$ is plotted for different $L$ and different $\rho$
(Fig. \ref{MTCparameters}). In the regime
of interest, i.e. for small reactivities, the profile is well described by the theory 
even for small $L$ and rather high reservoir densities.

\section{Conclusion}

Our simulations and analytical results describe the MTC effect
quantitatively over a wide range of parameters within the NBK model.
Moreover, trends which are independent of model details have
been identified by analytical calculations and lead to a more positive
result than concluded by \cite{Brau03} where no MTC effect at all
was reported for short interconnecting channels $L=1$. For reasonable
reactivities and channel lengths the MTC effect vanishes proportionally to
$1/N$, i.e., is inversely proportional to the grain diameter. This was
shown explicitly for a two-dimensional simulation model, but the
reasoning that led to this conclusion extends straightforwardly to a
three-dimensional system. Nevertheless, for optimized external 
process parameters
the NBK model exhibits an enhancement of the effective
reactivity of up to approx. $30\%$ for small grains and any
(even short) channel length and reactivity $c$. In the present 
study we have not taken into consideration that smaller molecules 
may diffuse into larger channels, not only the smaller channels. 
This was done in order to keep the model simple so that the origin 
of the observed MTC effect becomes as transparent as possible. We 
did make simulations for a different model where this mechanism is 
taken into account. We have found in these simulations that the MTC 
effect becomes stronger. In summary, our investigations suggest 
that MTC may enhance significantly the effective reactivity in zeolitic
nanoparticles with suitable binary channel systems and thus may
be of practical relevance in applications.

\end{document}